\documentclass[10pt,fleqn]{lhep}
\usepackage[T1]{fontenc}
\usepackage{amsmath}
\usepackage{amssymb} 
\usepackage[small]{caption} 
\usepackage{tikz} %
\newcommand{\tikznode}[2]{\tikz[baseline=(#1.base),remember picture]{\node (#1){$#2$};}}
\usetikzlibrary{positioning,shadows.blur,shapes.geometric,shapes.misc}
\tikzset{block/.style={on grid=true,rounded corners,align=center,blur
shadow,rectangle,draw,top color= white,bottom color=blue!15,minimum
height=4em,text width=2.8cm},
	yn/.style={on grid=true,rounded corners,align=center,blur
shadow,rectangle,midway,fill=yellow!15,draw,text width=0.54cm,text
depth=0pt},
	decision/.style={on grid=true,align=center,blur
shadow,regular polygon,chamfered rectangle,chamfered rectangle xsep=1.4ex,
draw,top color= white,bottom color=red!15},
    arrowFix1/.style={shorten <=-2pt},
    arrowFix2/.style={shorten >=-2pt}}
\usepackage{mathrsfs}	
\usepackage{cancel}
\usepackage{pifont} 
\usepackage{mathtools}
\usepackage{braket}
\usepackage{xfrac}
\usepackage{xspace}
\usepackage{units}
\usepackage{subfig}
\usepackage{cleveref}
\usepackage{soul}
\usepackage{xcolor}

\DeclareMathOperator{\ord}{ord}
\newcommand{\CenterObject}[1]{\ensuremath{\vcenter{\hbox{#1}}}}
\newcommand{\D}{\mathrm{d}}
\newcommand{\I}{\mathrm{i}}

\newcommand{\SO}[1]{\ensuremath{\mathrm{SO}(#1)}}

\newcommand{\SU}[1]{\ensuremath{\mathrm{SU}(#1)}}
\newcommand{\U}[1]{\ensuremath{\mathrm{U}(#1)}}
\newcommand{\Z}[1]{\ensuremath{\mathbb{Z}_{#1}}} 
\newcommand{\ChargeC}{\ensuremath{\mathcal{C}}\xspace}

\newcommand{\ParityP}{\ensuremath{\mathcal{P}}\xspace}
\newcommand{\CP}{\ensuremath{\mathcal{CP}}\xspace}
\newcommand{\DiscreteGroup}{\ensuremath{G}\xspace}
\newcommand{\UCP}{\ensuremath{\red{U_\mathrm{CP}}}\xspace}

\newcommand{\FSI}{\ensuremath{\mathrm{FS}_u}}

\newcommand{\UU}[2][]{\ensuremath{U_{\rep[#1]{#2}}}}

\newcommand{\rhoR}[1]{\ensuremath{\rho_{\rep[#1]{r}}}}
\newcommand{\chiR}[1]{\ensuremath{\chi_{\rep[#1]{r}}}}

\newcommand{\Tprime}{\ensuremath{\mathrm{T}'}}
\newcommand{\Afour}{\ensuremath{A_4}}

\newcommand{\CPgen}{\ensuremath{\boldsymbol{\widetilde{\ChargeC\ParityP}}}}

\newcommand{\SemiDirect}[0]{\ensuremath{\rtimes}}

\newcommand*{\rep}[2][]{\ensuremath{{\boldsymbol{#2}#1}}} 
\newcommand{\crep}[1]{\ensuremath{\overline{\boldsymbol{#1}}}}

\newcommand{\op}[1]{\ensuremath{\boldsymbol{#1}}}

\allowdisplaybreaks

\DeclareCollectionInstance{plainmath}{xfrac}{mathdefault}{math}
{
scale-factor = 0.9
}
\UseCollection{xfrac}{plainmath}

\begin{document}
\title{Group--theoretical origin of \CP violation}
\author{Mu--Chun Chen and Michael Ratz}
\address{University of California Irvine, Irvine CA 92697, USA}

\begin{abstract}
This is a short review of the proposal that \CP violation may be due to the fact
that certain finite groups do not admit a physical \CP transformation. This
origin of \CP violation is realized in explicit string compactifications
exhibiting the Standard Model spectrum.
\end{abstract}

\maketitle

\section{Introduction}

As is well known, the flavor sector of the Standard Model (SM) violates \CP, the
combination of the discrete symmetries $\mathcal{C}$ and $\mathcal{P}$. This
suggests  that flavor and \CP violation have a common
origin~\cite{Chen:2014tpa}. The question of flavor concerns the fact that the SM
fermions come in three families that are only distinguished by their masses. 
$\SU2_\mathrm{L}$ interactions lead to transitions between these families, which
are governed by the mixing parameters in the CKM and PMNS matrices.  These
mixing parameters are completely unexplained in the SM.  Furthermore, \CP
violation manifests in the SM through the the non--zero phase $\delta_{q}$ in
the CKM matrix.  In the lepton sector, the latest measurements from T2K as well
as the global fit for neutrino oscillation  parameters also hint at non--zero
value for the Dirac phase $\delta_{\ell}$ in the PMNS matrix, which will, if
proved, establish violation of \CP in the lepton sector. 

The observed repetition of families, i.e.\ the fact that the quarks and leptons
appear in 3 generations, hint at a flavor symmetry under which the generations
transform nontrivially. The main punchline of this review is the statement that
certain flavor symmetries clash with \CP. In other words, \CP violation can be
entirely group theoretical in origin.   

\subsection{What is a physical \CP transformation?}

Charge conjugation $\mathcal{C}$ inverts, by definition, all currents. This
implies that Standard Model representations $\rep{R}$ get mapped to their
conjugates, $\crep{R}$. Likewise, parity $\mathcal{P}$ exchanges the $(0,\nicefrac{1}{2})$
and $(\nicefrac{1}{2},0)$ representations of the Lorentz group, which
corresponds to complex conjugation at the level of $\mathrm{SL}(2,\mathbb{C})$.
That is, at the level of $G_\mathrm{SM}\times\mathrm{SL}(2,\mathbb{C})$ \CP is
represented by the (unique) nontrivial outer automorphism.

This fact has led to the suspicion that any nontrivial outer automorphism can
be used to coin a valid \CP transformation~\cite{Holthausen:2012dk}. However,
this turns out not to be the case~\cite{Chen:2014tpa}. To see this, let us
review why we care about whether or not \CP is violated. One reason we care is
that \CP violation is a prerequisite for baryogenesis~\cite{Sakharov:1967dj},
i.e.\ the creation of the matter--antimatter asymmetry of our universe.
Therefore, a physical \CP transformation exchanges particles and antiparticles,
a requirement an arbitrary outer automorphism may or may not fulfill. As
discussed in detail in~\cite{Chen:2014tpa}, \CP transformations are linked to
\emph{class--inverting} outer automorphisms. 

\subsection{\CP and Clebsch--Gordan coefficients}

It turns out that some finite groups do not have such outer automorphisms but
still complex representations. These groups thus clash with \CP! Further, they
have no basis in which all Clebsch--Gordan coefficients (CGs) are real, and \CP
violation can thus be linked to the complexity of the CGs~\cite{Chen:2009gf}. 

\section{\boldmath\CP violation from finite groups\unboldmath}

\subsection{The canonical \CP\ transformation}

Let us start by collecting some basic facts. Consider a scalar field operator
\begin{equation}
 \op{\phi}(x)
 ~=~
 \int\!\D^3p\,\frac{1}{2E_{\vec p}}\,\left[
 \op{a}(\vec p)\,\mathrm{e}^{-\I\,p\cdot x}
 +
 \op{b}^\dagger(\vec p)\,\mathrm{e}^{\I\,p\cdot x}
 \right]\;,
\end{equation} 
where \op{a} annihilates a particle and $\op{b}^\dagger$ creates an
antiparticle. The \CP operation exchanges particles and antiparticles,
\begin{subequations}
\begin{align}
 \left(\op{\ChargeC}\,\op{\ParityP}\right)^{-1}\,\op{a}(\vec p)\,\op{\ChargeC}
 \,\op{\ParityP}
 &~=~\eta_{\ChargeC\ParityP}\,\op{b}(-\vec p)\;,\\
 \left(\op{\ChargeC}\,\op{\ParityP}\right)^{-1}\,\op{a}^\dagger(\vec p)\,\op{\ChargeC}
 \,\op{\ParityP}
 &~=~\eta_{\ChargeC\ParityP}^*\,\op{b}^\dagger(-\vec p)
 \\
 \left(\op{\ChargeC}\,\op{\ParityP}\right)^{-1}\,\op{b}(\vec p)\,\op{\ChargeC}
 \,\op{\ParityP}
 &~=~\eta_{\ChargeC\ParityP}^*\,\op{a}(-\vec p)\;,\\
 \left(\op{\ChargeC}\,\op{\ParityP}\right)^{-1}\,\op{b}^\dagger(\vec p)\,\op{\ChargeC}
 \,\op{\ParityP}
 &~=~\eta_{\ChargeC\ParityP}\,\op{a}^\dagger(-\vec p)\;,
\end{align} 
\end{subequations}
where $\eta_{\CP}$ is a phase factor. On the scalar fields, \CP transformations 
act as
\begin{align}
\phi(x)~\xmapsto{\op{\ChargeC}\,\op{\ParityP}}~
\eta_{\ChargeC\ParityP}\,\phi^*(\mathcal{P}x)\;.
\end{align}
At this level, $\eta_{\CP}$ can be viewed as the freedom of rephasing the field,
i.e.\ a choice of field basis. Later, when we replace $\eta_{\CP}$ by some
matrix $U_{\CP}$, this will still reflect the freedom to choose a basis. The
important message here is that there is a well--defined operation, the \CP
transformation, which exchanges particles with antiparticles. It is this very
transformation which is broken in the $K^0-\overline{K^0}$ system, and whose
violation is a prerequisite for baryogenesis. 

\subsection{\CP\ vs.\ outer automorphisms}

\begin{figure}[htb]
\centering
\includegraphics{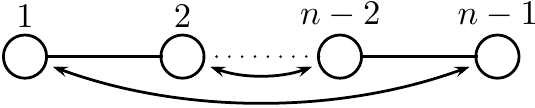}
\caption{\CP acts as the unique nontrivial outer automorphism on the \SU{N}
groups.}
\label{fig:OutSUN}
\end{figure}

Next let us review what \CP does in the context of  most of the continuous
(i.e.\ Lie) groups. If the representation under consideration is real, the
canonical \CP\ does the job. For complex representations, \CP involves a
nontrivial outer automorphism (cf.~\Cref{fig:OutSUN}).

In particular, in the context of the standard model gauge group and the usual
theories of grand unification (GUTs),
\[
G_\mathrm{SM}=\SU3_\mathrm{C}\times\SU2_\mathrm{L}\times\U1_Y\subset\SU5\subset\SO{10}\subset\text{E}_6
\]
\CP\ always involves outer automorphisms,
\begin{align*}
\CenterObject{\includegraphics{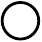}}
\times\CenterObject{\includegraphics{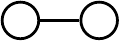}}&~\subset~
\CenterObject{\includegraphics{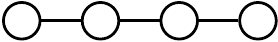}}\\
&~\subset~\CenterObject{\includegraphics{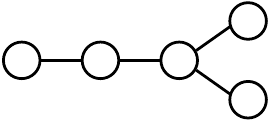}}\\
&~\subset~\CenterObject{\includegraphics{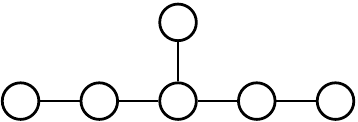}}
\end{align*}

One may thus expect that this also true for discrete (family) symmetries.
However, this is an accident, and is already not the case for \SO8, the only Lie
group with a non--Abelian outer automorphism group, namely $S_3$. Which of those
outer automorphisms, if any, corresponds to the physical \CP transformation? As we
shall discuss next, in particular for finite it is not true that there is a
unique outer automorphism, nor that any nontrivial outer automorphism qualifies
as a physical \CP transformation~\cite{Chen:2009gf,Chen:2014tpa}.

\subsection{Generalized \CP\ transformations}
\label{sec:GeneralizingCP}

To see this, consider a setting with discrete symmetry \DiscreteGroup.
One can now impose a so--called \emph{generalized \CP transformation},
\begin{subequations}\label{eq:GeneralizedCPOP}
\begin{align}
 \left(\op{\ChargeC}
 \,\op{\ParityP}\right)^{-1}\,\op{a}(\vec p)\,\op{\ChargeC}
 \,\op{\ParityP}
 &~=~\UCP\,\op{b}(-\vec p)\;,\\
 \left(\op{\ChargeC}\,\op{\ParityP}\right)^{-1}\,\op{a}^\dagger
 (\vec p)\,\op{\ChargeC}
 \,\op{\ParityP}
 &~=~\op{b}^\dagger(-\vec p)\,\UCP^\dagger
 \\
 \left(\op{\ChargeC}\,\op{\ParityP}\right)^{-1}\,\op{b}(\vec p)\,\op{\ChargeC}
 \,\op{\ParityP}
 &~=~\op{a}(-\vec p)\,\UCP^\dagger\;,
 \\
\left(\op{\ChargeC}\,\op{\ParityP}\right)^{-1}\,\op{b}^\dagger(\vec p)\,\op{\ChargeC}
 \,\op{\ParityP}
 &~=~\UCP\,\op{a}^\dagger(-\vec p)\;,
\end{align} 
\end{subequations}
where $\op{a}$ is a vector of annihilation operators and $\op{a}^\dagger$ is a
vector of creation operators. $\UCP$ is a unitary matrix.

The reader may wonder whether or not the need to ``generalize'' is specific to
the \CP transformation. This is not the case. A very close analogy is the
Majorana condition. In the Majorana basis, it boils down to the requirement that
$\Psi=\Psi^*$ for a Dirac spinor $\Psi$. However, in the Weyl or Dirac basis,
this condition becomes $\Psi=\mathsf{C}\,\Psi^*$ with some appropriate matrix
$\mathsf{C}$. That is, the antiparticle of a particle described by $\Psi$ is
described by $\mathsf{C}\,\Psi^*$, and not just $\Psi^*$. Likewise, in the above
discussion around \eqref{eq:GeneralizedCPOP}, the \CP conjugate  (i.e.\
antiparticle up to a transformation of the spatial coordinates) of a scalar
described by $\phi$ will be described by $U_{\CP}\,\phi^*$, see
\eqref{eq:GeneralizedCP} below. So, in a way, $U_{\CP}$ is the analogy of the
matrix $\mathsf{C}$ for Dirac fermions. 

As is evident from this argument and as pointed out in \cite{Holthausen:2012dk},
generalizing \CP may not be an option, but a necessity. To see this, consider a
model in which $\DiscreteGroup$ is \Afour\ (or \Tprime). Then a
\Tprime--invariant contraction/coupling is given by
\begin{align}
 \left[\phi_{\rep[_2]{1}}\otimes
 \left(x_{\rep{3}}\otimes y_{\rep{3}}\right)_{\rep[_1]{1}}\right]_{\rep[_0]{1}} 
 & ~\propto~ \phi\,\left(
 x_{1}\,y_{1}+\omega^2\,x_{2}\,y_{2}+\omega\,x_{3}\,y_{3}\right)\;,
\end{align}
where $\omega=\mathrm{e}^{2\pi\,\I/3}$. Crucially, the  canonical \CP
transformation maps this invariant  contraction to something noninvariant,
\begin{align}\label{eq:CanonicalCPTprime}
 x~\xmapsto{\op{\mathcal{C}}\,\op{\mathcal{P}}}~x^*
 \quad\&\quad
 y~\xmapsto{\op{\mathcal{C}}\,\op{\mathcal{P}}}~y^*\quad\&
 \quad
 \phi~\xmapsto{\op{\mathcal{C}}\,\op{\mathcal{P}}}~\phi^*\;.
\end{align}
Hence, the canonical \CP transformation is not an (outer) automorphism of
\Tprime (in this basis). Therefore, in order to warrant \CP conservation, one
needs to impose a so--called \emph{generalized \CP\ transformation} $\CPgen$
under which $\phi~\xmapsto{\CPgen}~\phi^*$ as usual but
\begin{align}
\begin{pmatrix}x_1\\ 
\tikznode{x2}{x_2}
\\ \tikznode{x3}{x_3}\end{pmatrix}
 ~\xmapsto{\red{\CPgen}}~
 \begin{pmatrix} x_1^*\\ \tikznode{x3s}{x_3^*}\\ 
\tikznode{x2s}{x_2^*}
 \end{pmatrix}
\begin{tikzpicture}[overlay,remember picture]
\draw[->,red,thick] (x2) edge (x2s);
\draw[->,red,thick] (x3) edge (x3s);
\end{tikzpicture}\;,
\quad
 \begin{pmatrix}y_1\\ 
\tikznode{y2}{y_2}
\\ \tikznode{y3}{y_3}\end{pmatrix}
 ~\xmapsto{\red{\CPgen}}~
 \begin{pmatrix}y_1^*\\ \tikznode{y3s}{y_3^*}\\ 
\tikznode{y2s}{y_2^*}
 \end{pmatrix}\;.
\begin{tikzpicture}[overlay,remember picture]
\draw[->,red,thick] (y2) edge (y2s);
\draw[->,red,thick] (y3) edge (y3s);
\end{tikzpicture}
\end{align}

\subsection{Constraints on generalized \CP transformations}

In order for a \CP transformation not to clash with the group, i.e.\ in order to
avoid mapping something that is invariant under the symmetry transformations to
something that isn't (cf. \eqref{eq:CanonicalCPTprime}), it has to be an
automorphism $\red{u}~:~\DiscreteGroup \to\DiscreteGroup$ of the group. 
An automorphism $u$ corresponding to a physical \CP transformation has to
fulfill the consistency condition~\cite{Holthausen:2012dk} (see
also~\cite{Feruglio:2012cw})
\begin{align}
\rho\bigl(u(g)\bigr)
~=~
\UCP\,\rho(g)^*\,\UCP^{\dagger} 
 \quad\forall~g\in\DiscreteGroup\;.
\end{align}

Here, \UCP is a unitary matrix that enters the generalized \CP transformation,
\begin{align}\label{eq:GeneralizedCP}
 \Phi(x)~\xmapsto{\CPgen}~U_\mathrm{CP}\,
 \Phi^*(\mathcal{P}\,x)\;,
\end{align}
where $\Phi$ denotes collectively the fields of the theory/model, and
$\mathcal{P}\,(t,\vec x)=(t,-\vec x)$ as usual. In particular, each
representation gets mapped on its own conjugate, i.e.\ $U_\mathrm{CP}$ is
block--diagonal in \Cref{eq:GeneralizedCP}, 
\begin{align}
 \begin{pmatrix}\uparrow\\ \phi_{\rep[_{i_1}]{r}}\\ \downarrow \\ \hline 
 \uparrow\\ \phi_{\rep[_{i_2}]{r}}\\ \downarrow \\ \hline \vdots
 \end{pmatrix}
 ~\xmapsto{\red{\CPgen}}~
 \left(\begin{array}{@{}ccc|ccc|c@{}}
 \nwarrow & & \nearrow & & & & \\
 &\!\!\! \UU[_{i_1}]{r} \!\!\!&  & & & & \\
 \swarrow & & \searrow & & & & \\
 \hline
 & & & \nwarrow  & & \nearrow &  \\
 & & & & \!\!\!\UU[_{i_2}]{r}\!\!\!  & & \\
 & & & \swarrow  & & \searrow &  \\
 \hline 
 & & & & & & \ddots
 \end{array}\right)
 \,
 \begin{pmatrix}\uparrow\\ \phi_{\rep[_{i_1}]{r}}^*\\ \downarrow \\ \hline 
 \uparrow\\ \phi_{\rep[_{i_2}]{r}}^*\\ \downarrow \\ \hline \vdots
 \end{pmatrix}\;,\label{eq:CPMatrixTrafo}
\end{align}
where the $\UU[_{i_a}]{r}$ are unitary matrices that depend on the
representation $\rep{r}_{i}$ only. The $a$ subscripts in
$\phi_{\rep[_{i_a}]{r}}$ label the particles whereas the $i_a$ subscripts
indicate the representations, i.e.\ different particles can furnish the same
representations under \DiscreteGroup. The transformation law
\eqref{eq:CPMatrixTrafo} disagrees with \cite{Holthausen:2012dk}, where it was
suggested that one can use any outer automorphism in order to define a viable
\CP transformation. 

Therefore, the requirement that the candidate transformation is a
\emph{physical} \CP transformation, which exchanges particles and their
antiparticles, amounts to demanding that $u$ be class--inverting. In all known
cases, $u$ can be taken to be an automorphism of order two. Of course, this does
not exclude the interesting possibility to make \CP part of a higher--order
transformation~\cite{Ivanov:2015mwl}. 

\subsection{\CP vs.\ \CP--like transformations}

However, it is important to distinguish physical \CP transformations, and their
\emph{proper} generalizations, from \CP--like transformations. Unfortunately,
the latter have sometimes been called ``generalized \CP transformations'' in the
literature. However, some of the proposed ``generalized \CP transformations'' do
not warrant physical \CP conservation. Thus they do not have a connection to the
observed \CP violation in the CKM sector, nor to baryogenesis and so on. That
is, the violation of physical \CP is a prerequisite of a nontrivial decay
asymmetry, but the violation of a so--called ``generalized \CP transformation''
is not. That is to say some of the operations dubbed ``generalized \CP
transformation'' in the literature are not physical \CP transformations, which
is why we refer to them as ``\CP--like''.

Given all these considerations, it is a valid question whether or not one can
impose a \emph{physical} \CP in any model. As mentioned above, this is not the
case. Certain finite symmetries clash with \CP. Here ``clash'' means that any
physical \CP transformation maps some \DiscreteGroup--invariant term(s) on
non-invariant terms, and thus does not comply with \DiscreteGroup, i.e.\ is not
an automorphism thereof.  We will discuss next how one may tell those symmetries
that clash with \CP apart from those which do not.

\subsection{The Bickerstaff--Damhus automorphism (BDA)}

In a more group--theoretical language the question whether or not one can impose
\CP can be rephrased as whether or not a given finite group has a so--called
Bickerstaff--Damhus automorphism (BDA)~\cite{Bickerstaff:1985jc} $u$, 
\begin{align}
 \rhoR{_i}\!\left(u(g)\right)
 &~=~\UU[_i]{\rep{r}}\,\rhoR{_i}\!(g)^*\,
 \UU[_i]{\rep{r}}^\dagger\quad\forall~g\in\DiscreteGroup
 ~\text{and}~\forall~i\;,
\end{align}
where $U$ is unitary and symmetric. The existence of a BDA implies the existence
of a basis in which all Clebsch--Gordan (CG) coefficients are real. In physics,
such a basis is often referred to as ``\CP basis''. The connection between the
BDA, the complexity of the CG's, and \CP has first been pointed out in
\cite{Chen:2009gf}.\footnote{However, the example used there, $\Tprime$, turns out
not to be of the \CP violating type.}

Of course, this raises the question whether or not one can tell if a given group
has an BDA. There is a rather simple criterion for this, based on the so--called
extended twisted Frobenius--Schur indicator (see
\cite{Bickerstaff:1985jc,Kawanaka1990} for the so--called twisted
Frobenius--Schur indicator),
\begin{align}
  \mathrm{FS}^{(n)}_{u}(\rep[_i]{r}) ~:=~ \frac{(\dim{\rep[_i]{r}})^{n-1}}{|\DiscreteGroup|^n}\,
  \sum_{g_i \in \DiscreteGroup} \, \chiR{_i}(g_1 \, u(g_1)\cdots g_n
  \, u(g_n))\;,
\end{align}
where $\chiR{_i}$ denotes the character and
\begin{align}
 n&~=~\begin{cases}
 \ord({u})/2 &~\text{if $\ord({\blue{u}})$ is even,}\\
 \ord({u}) &~\text{if $\ord({\blue{u}})$ is odd.}
 \end{cases}
\end{align}
It has the crucial property
\begin{align}
  \mathrm{FS}^{(n)}_{\blue{u}}(\rep[_i]{r}) ~=~ 
  \pm1 ~ \forall~i\quad\Longleftrightarrow\quad
  u\text{~is class--inverting}\;.
\end{align}
So one has to scan over all candidate automorphism $u$ to determine whether one
of them is a BDA, a task that can be automatized.
\begin{figure*}[!htb]
\centering
\begin{tikzpicture}[thick,node distance=5.7cm,>=stealth]
    \node[decision] (odd) at (0,0) {order $|\DiscreteGroup|$\\ of \DiscreteGroup\ is odd};
    \node[decision,right=of odd] (involutoryCI) {\DiscreteGroup\ has\\class--inverting\\ involutory\\ automorphisms};
    \node[decision,right=of involutoryCI] (oddirreps) {\DiscreteGroup\ has only\\ irreps of odd\\ dimension};
    \node[decision,below=5cm of involutoryCI] (equation) {there is an\\ automorphism $u$\\ with all \FSI's\\ equal to $+1$};
    \draw[->,ultra thick] (odd.east) -- (involutoryCI.west) node[yn] {no};
    \draw[->,ultra thick] (involutoryCI) -- (oddirreps) node[yn] {yes};
    \draw[->,ultra thick] (oddirreps.-150) -- (equation.30) node[yn] {no};
    \node[block,below=5cm of oddirreps] (yes) {there exists a basis\\ with real CG's};
    \node[block,below=5cm of odd] (no) {there exists no basis\\ with real CG's};
    \draw[->,ultra thick] (odd.south) -- (no.north) node[yn] {yes};
    \draw[->,ultra thick,arrowFix2] (involutoryCI.south west) -- (no.north east) node[yn] {no};
    \draw[->,ultra thick] (oddirreps.south) -- (yes) node[yn] {yes};
    \draw[->,ultra thick] (equation.west) -- (no) node[yn] {no};
    \draw[->,ultra thick] (equation.east) -- (yes) node[yn] {yes};
  \end{tikzpicture}
\caption{Sequence of steps to determine whether or not a group admits a basis in
which all CG's are real. From \cite{Chen:2014tpa}.}
\label{fig:RealCGs}
\end{figure*}
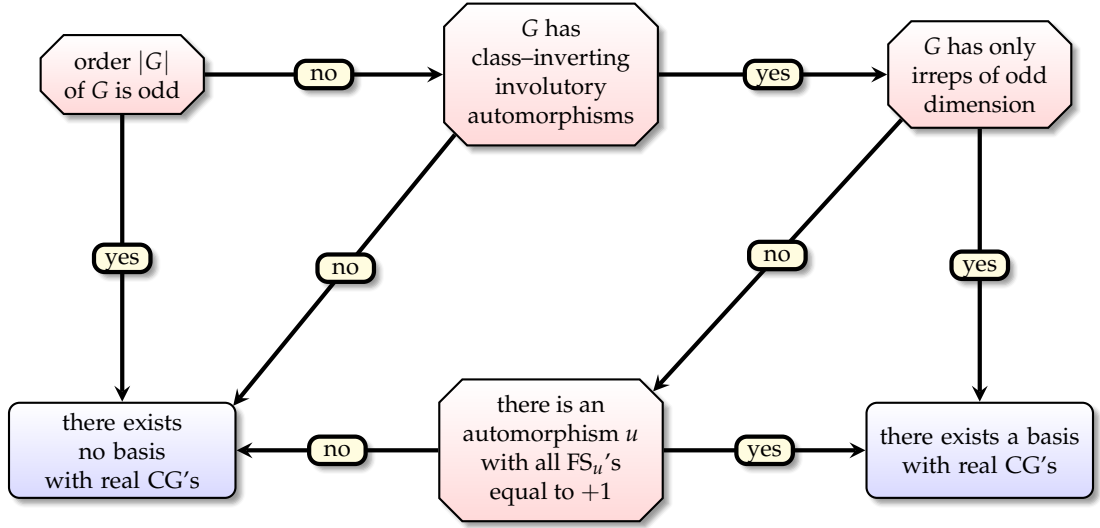

Even though the steps in Figure~\ref{fig:RealCGs} may, at first sight, appear a
bit cumbersome, one should remember that they allow us to uniquely determine, in
an automatized way, whether or not a symmetry has a basis in which all CG's are
real, or, if a symmetry clashes with \CP. Of course, this analysis is
independent of bases, as it should be.

\subsection{Three types of groups}

\begin{figure*}[!h!]
\centering
\begin{tikzpicture}[thick,node distance=5.5cm,>=stealth]
    \node[block] (init) at (0,0) {group \DiscreteGroup\ with automorphisms $u$}; 
    \node[decision,right=of init] (ci) {there is a
        \\$u$ for which\\ no $\FSI^{(n)}$ is $0$};
    \node[block,right=of ci] (typeII) {\textbf{Type II}: $u$ defines a physical
	\CP transformation};
    \node[decision,below=7em of typeII] (FSIpm) {there is an
     \\ involutory $u$
     \\ for which all
     \\ $\FSI^{(1)}$ are $+1$};
    \node[block,below=10em of ci.south] (typeIIA) {\textbf{Type II~A}: there is
	a \CP basis in which all CG's are real};
    \node[block,right=of typeIIA.south,anchor=south] (typeIIB) {\textbf{Type II~B}: there is no basis in which all CG's are real};
    \node[block,left=of typeIIA.south,anchor=south] (typeI) {\textbf{Type I}:
	generic settings based on $\DiscreteGroup$ do not allow for a physical \CP transformation};
    \draw[->,ultra thick] (init) -- (ci);
    \draw[->,ultra thick] (ci.south west) |-+(0,-2.5em)-| (typeI.north) node[yn] {no};
    \draw[->,ultra thick] (ci) -- (typeII) node[yn] {yes};
    \draw[->,ultra thick] (typeII) -- (FSIpm);
    \draw[->,ultra thick] (FSIpm.west) -| (typeIIA) node[yn] {yes};    
    \draw[->,ultra thick] (FSIpm.south) -- (typeIIB) node[yn,yshift=0.25em] {no};   
  \end{tikzpicture}
  \caption{The regular and extended twisted
   Frobenius--Schur indicators \FSI\ and $\FSI^{(n)}$ allow us to distinguish
   between the three types of groups. Here, $n$ is $n=\ord(u)/2$ for even
and $n=\ord(u)$ for odd $\ord(u)$. From \cite{Chen:2014tpa}.}
  \label{fig:3types} 
\end{figure*}
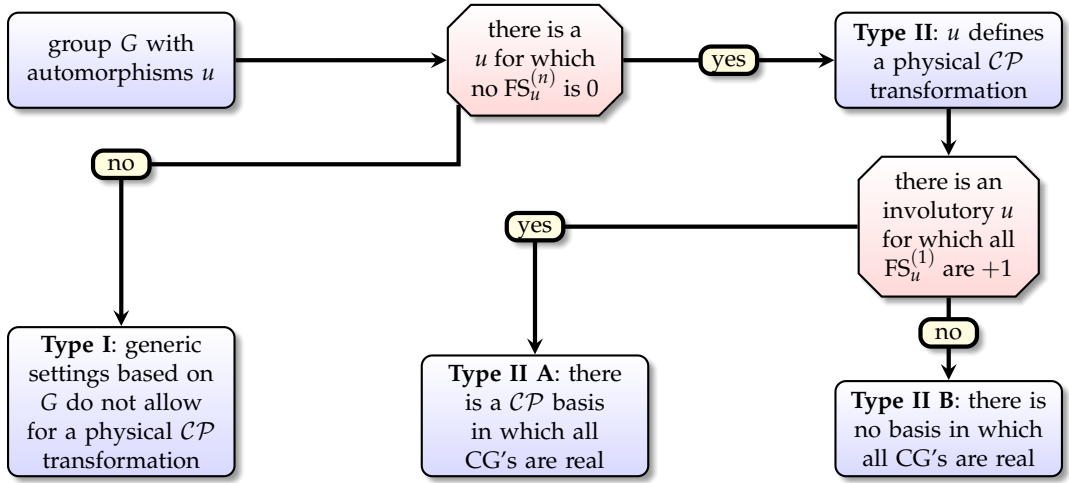

Given these tools, one can distinguish between three types of groups
\cite{Chen:2014tpa}:
\begin{description}
 \item[Case~I:] for all involutory automorphisms $u_\alpha$ of 
 the flavor group there is at least one representation $\rep[_i]{r}$ for which
 $\text{FS}_{u_\alpha}(\rep[_i]{r})=0$. Such discrete symmetries clash with \CP.
 
 \item[Case~II:] there exists an involutory automorphism $u$ for which
 the \FSI's for all representations are non--zero.  Then
 there are two sub--cases:
 \begin{description}
  \item[Case~II~A:] all \FSI's are $+1$ for one of those $u$'s.
  In this case, there exists a basis with real Clebsch--Gordan coefficients. 
  The BDA is then the automorphism of the physical \CP\ transformation/
  \item[Case~II~B:] some of the \FSI's are $-1$ for all candidate $u$'s. That
  means that there exists no BDA, and, as a consequence, one cannot find a basis in which all CG's are
  real. Nevertheless, any of the $u$'s can be used to define a \CP
  transformation.  
 \end{description}
\end{description}
The distinction between the groups is illustrated in
Figure~\ref{fig:3types}.

\section{\boldmath\CP violation with an unbroken \CP transformation\unboldmath}

Having seen that there are finite groups that do not admit a physical \CP
transformation, one may wonder about the following question: if one obtains this
finite group from a continuous one by spontaneous breaking, at which stage does
\CP violation arise? That is, take an \SU{N} gauge symmetry, impose \CP, and
break it down to a type--I subgroup. The obvious options how \CP violation may
come about include
\begin{enumerate}
 \item \CP gets broken by the VEV that breaks \SU{N} to \DiscreteGroup and
 \item the resulting setting always has additional symmetries and does not
 violate \CP.
\end{enumerate}
Rather surprisingly, none of these is the true answer. As demonstrated
explicitly in an example in which an \SU3 symmetry gets broken to
$T_7=\Z3\SemiDirect\Z7$, the outer automorphism of \SU3 merges into the outer
automorphism of $T_7$, which however does \emph{not} entail \CP conservation
\cite{Ratz:2016scn}. 

This leads to a novel way to address the strong \CP problem. Start with a
theory based on $\SU3_\mathrm{C}\times\SU3_\mathrm{F}$ (and of course the other
gauge symmetries of the standard model). Now impose \CP, which implies that the
coefficient $\theta$ of the QCD $G_{\mu\nu}\widetilde{G}^{\mu\nu}$ term vanishes. Next
break the continuous flavor symmetry down to a type I flavor symmetry. Then
$\theta$ still vanishes, but \CP is violated in the flavor sector. This is what
is required to solve the strong \CP problem of the standard model. An explicit
example will be discussed elsewhere.

\section{\boldmath\CP violation from strings\unboldmath}

Of course, there are alternatives to embedding the discrete flavor symmetry
\DiscreteGroup into a continuous gauge symmetry (in four spacetime dimensions).
In fact, anomaly considerations seem to disfavor this possibility: an
$\SU3_\mathrm{F}$ symmetry with the families transforming as \rep{3}--plets has
un-cancelled anomalies (see e.g.~\cite{Zwicky:2009vt}). On the other hand,
non--Abelian discrete flavor symmetries may originate from extra
dimensions~\cite{Altarelli:2006kg}. In particular, orbifold compactifications of
the heterotic string lead to various flavor
groups~\cite{Kobayashi:2006wq,Olguin-Trejo:2018wpw}. These symmetries originate
from gauge symmetries in higher dimensions~\cite{Beye:2014nxa,Beye:2015wka}, as
they should~\cite{Witten:2017hdv}. As it turns out, already the very first
3--generation orbifold model~\cite{Ibanez:1987sn} has a $\Delta(54)$ flavor
symmetry~\cite{Kobayashi:2006wq}, which is according to the
classification~\cite{Chen:2014tpa} type I and thus \CP--violating. Therefore,
\CP is violated in such models~\cite{Nilles:2018wex}.

\begin{figure*}
\centering
\subfloat[]{\label{fig:WindingModes1}
\includegraphics[scale=0.6]{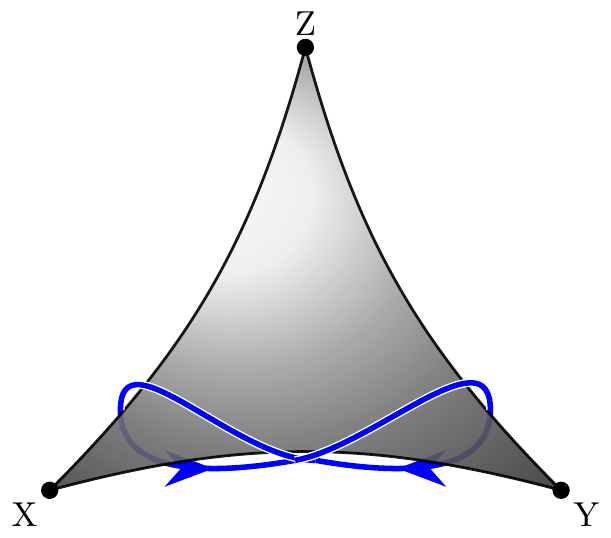}}
~
\subfloat[]{\label{fig:WindingModes2}
\includegraphics[scale=0.6]{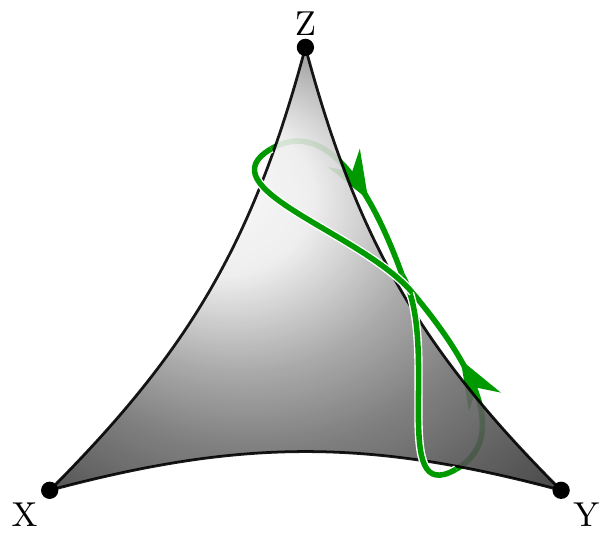}}
~
\subfloat[]{\label{fig:WindingModes3}
\includegraphics[scale=0.6]{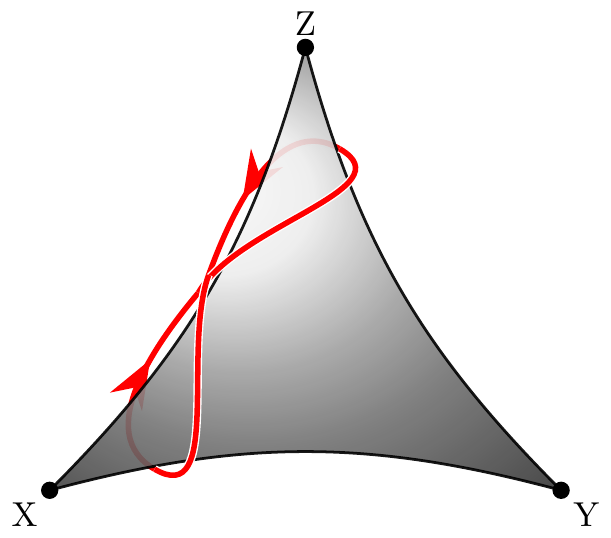}}
~
\subfloat[]{\label{fig:WindingModes4}
\includegraphics[scale=0.6]{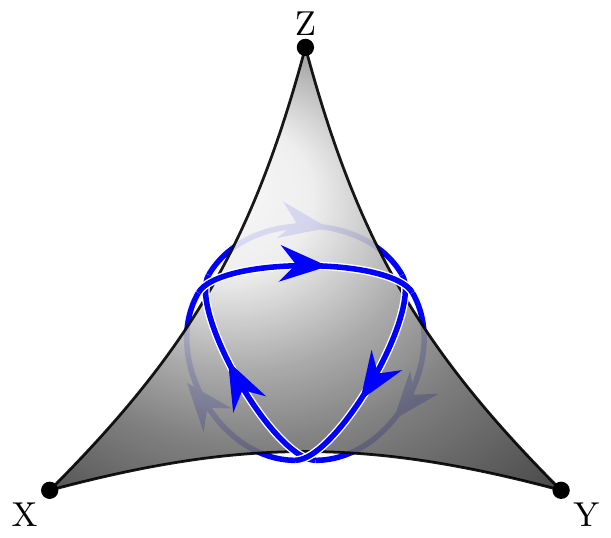}}
\caption{Winding strings. Linear combinations of \ref{fig:WindingModes1}--\ref{fig:WindingModes3}
give rise to three $\Delta(54)$ doublet representations while
\ref{fig:WindingModes4} leads to the fourth and last one.}
\label{fig:WindingStrings}
\end{figure*}

When establishing explicitly that \CP is violated, it was noticed that at the
massless level only 1- and 3--dimensional representations of $\Delta(54)$ occur.
There exist outer automorphisms of $\Delta(54)$ which map all these
representations on their conjugates. However, this is no longer the case when
one includes the massive states. In particular, the winding strings (see
\Cref{fig:WindingStrings}) give rise to $\Delta(54)$ doublets.

The presence of these doublets leads to \CP violation~\cite{Nilles:2018wex}.
This can be made explicit by finding a basis--invariant contraction (see
\cite{Bernabeu:1986fc}) that has a nontrivial phase. Of course, at this level
the flavor symmetry is unbroken, and there is no direct connection between the
phase of the contraction presented in~\cite{Nilles:2018wex} and the \CP
violation in the CKM matrix or baryogenesis. One would have to study explicit
models (e.g.~\cite{Carballo-Perez:2016ooy}) in which the flavor symmetry gets
broken and potentially realistic mass matrices arise how this \CP violation from
strings manifests itself in the low--energy effective theory. This has not yet
been carried out. Nevertheless, it is clear that if \CP is broken at the
orbifold point, it won't un-break by moving away from it by e.g.\ giving the
flavons VEVs. 

More recently, an additional amusing observation has been
made~\cite{Baur:2019kwi}. In orbifold compactifications (without the so--called
Wilson lines~\cite{Ibanez:1986tp}), the flavor symmetry \DiscreteGroup is simply
the outer automorphism group of the space group $\mathbb{S}$. In a bit more
detail, the states of an orbifold correspond to conjugacy classes of the space
group, and can be represented by space group elements $(\theta^k,n_\alpha
e_\alpha)$, where $\theta^k$ stands for a discrete rotation and $n_\alpha
e_\alpha$ an element of the underlying torus lattice. These conjugacy classes
form multiplets under the outer automorphism group of the space group, thus
$G=\text{out}(\mathbb{S})$. This leads to the picture of ``out of out'',
\begin{equation}
 \CP~\in~\text{out}(G)~=~\text{out}\bigl(\text{out}(\mathbb{S})\bigr)\;.
\end{equation}

Let us also mention that other orbifold geometries come with different flavor
symmetries. The probably simplest option is a $\mathbb{Z}_2$ orbifold plane,
which leads to a $D_4$ family symmetry~\cite{Kobayashi:2004ya,Kobayashi:2006wq}.
$D_4$ is a type II group, meaning that here one cannot immediately conclude that
\CP is violated. On the other hand, it entails a $\Z4^R$ symmetry, which solves
several shortcomings of the supersymmetric standard model at 
once~\cite{Babu:2002tx,Lee:2010gv,Lee:2011dya,Kappl:2010yu}. In particular, it
solves the $\mu$ problem and explains the longevity of the proton and the
stability of the LSP. All these examples illustrate the impact of properties of
compact dimensions on particle phenomenology. 

Arguably, it is rather amusing that \CP violation can be tied to the presence of
states that are required anyway for completing the models in the ultraviolet.
One may thus say that, at least in these models, consistency in the ultraviolet
requires \CP to be violated.

\section{Summary}

\CP violation may originate from group theory. We have reviewed the observation
that there are certain finite groups that clash with \CP in the sense that, if
these groups are realized as (flavor) symmetries, \CP is violated. To the best
of our knowledge, this is a situation that is not too ubiquitous in theory space.
What usually happens is that an extra symmetry results from imposing a symmetry.
Here, the opposite happens: \CP can get broken because another (flavor) symmetry
is imposed or emerges.

These \CP--breaking symmetries emerge from explicit string models. Even the
earliest 3--generation string models in the literature have a \CP violating
discrete symmetry. In the string models, all symmetries have a clear geometric
interpretation, which is why it is fair to say that the origin of \CP violation
described in this review deserves to be called ``geometric''.

\subsection*{Acknowledgements}

We are indebted to Dillon Berger for valuable comments on this review. We would
like to thank Ernest Ma for pushing us to write this up, and the Valencia group
for inviting us to their beautiful city. The work of M.C.C. was supported by, in
part, by the National Science Foundation under Grant No.\ PHY-1620638. The work
of M.R.\ is supported by NSF Grant No.\ PHY-1719438.

\bibliography{Orbifold}

\providecommand{\bysame}{\leavevmode\hbox to3em{\hrulefill}\thinspace}
\frenchspacing
\newcommand{\origttfamily}{}
\let\origttfamily=\ttfamily
\renewcommand{\ttfamily}{\origttfamily \hyphenchar\font=`\-}

\begin{thebibliography}{10}

\bibitem{Chen:2014tpa}
M.-C. Chen, M.~Fallbacher, K.~Mahanthappa, M.~Ratz, and A.~Trautner, Nucl.
  Phys. \textbf{B883} (2014), 267, \texttt{arXiv:1402.0507} [hep-ph].

\bibitem{Holthausen:2012dk}
M.~Holthausen, M.~Lindner, and M.~A. Schmidt, JHEP \textbf{1304} (2013), 122,
  \texttt{arXiv:1211.6953} [hep-ph].

\bibitem{Sakharov:1967dj}
A.~Sakharov, Pisma Zh.Eksp.Teor.Fiz. \textbf{5} (1967), 32.

\bibitem{Chen:2009gf}
M.-C. Chen and K.~Mahanthappa, Phys. Lett. \textbf{B681} (2009), 444,
  \texttt{arXiv:0904.1721} [hep-ph].

\bibitem{Feruglio:2012cw}
F.~Feruglio, C.~Hagedorn, and R.~Ziegler, JHEP \textbf{1307} (2013), 027,
  \texttt{arXiv:1211.5560} [hep-ph].

\bibitem{Ivanov:2015mwl}
I.~P. Ivanov and J.~P. Silva, Phys. Rev. \textbf{D93} (2016), no.~9, 095014,
  \texttt{arXiv:1512.09276} [hep-ph].

\bibitem{Bickerstaff:1985jc}
R.~Bickerstaff and T.~Damhus, International Journal of Quantum Chemistry
  \textbf{XXVII} (1985), 381.

\bibitem{Kawanaka1990}
N.~Kawanaka and H.~Matsuyama, Hokkaido Math.J. \textbf{19} (1990), 495.

\bibitem{Ratz:2016scn}
M.~Ratz and A.~Trautner, JHEP \textbf{02} (2017), 103,
  \texttt{arXiv:1612.08984} [hep-ph].

\bibitem{Zwicky:2009vt}
R.~Zwicky and T.~Fischbacher, Phys. Rev. \textbf{D80} (2009), 076009,
  \texttt{arXiv:0908.4182} [hep-ph].

\bibitem{Altarelli:2006kg}
G.~Altarelli, F.~Feruglio, and Y.~Lin, Nucl. Phys. \textbf{B775} (2007), 31,
  \texttt{arXiv:hep-ph/0610165} [hep-ph].

\bibitem{Kobayashi:2006wq}
T.~Kobayashi, H.~P. Nilles, F.~Pl{\"o}ger, S.~Raby, and M.~Ratz, Nucl. Phys.
  \textbf{B768} (2007), 135, \texttt{hep-ph/0611020}.

\bibitem{Olguin-Trejo:2018wpw}
Y.~Olguin-Trejo, R.~P{\'e}rez-Mart{\'i}nez, and S.~Ramos-S{\'a}nchez,
  \texttt{arXiv:1808.06622} [hep-th].

\bibitem{Beye:2014nxa}
F.~Beye, T.~Kobayashi, and S.~Kuwakino, Phys. Lett. \textbf{B736} (2014), 433,
  \texttt{arXiv:1406.4660} [hep-th].

\bibitem{Beye:2015wka}
F.~Beye, T.~Kobayashi, and S.~Kuwakino, JHEP \textbf{03} (2015), 153,
  \texttt{arXiv:1502.00789} [hep-ph].

\bibitem{Witten:2017hdv}
E.~Witten, Nature Phys. \textbf{14} (2018), 116, \texttt{arXiv:1710.01791}
  [hep-th].

\bibitem{Ibanez:1987sn}
L.~E. Ib{\'a}{\~n}ez, J.~E. Kim, H.~P. Nilles, and F.~Quevedo, Phys. Lett.
  \textbf{B191} (1987), 282.

\bibitem{Nilles:2018wex}
H.~P. Nilles, M.~Ratz, A.~Trautner, and P.~K.~S. Vaudrevange,
  \texttt{arXiv:1808.07060} [hep-th].

\bibitem{Bernabeu:1986fc}
J.~Bernab\'{e}u, G.~Branco, and M.~Gronau, Phys. Lett. \textbf{B169} (1986),
  243.

\bibitem{Carballo-Perez:2016ooy}
B.~Carballo-Perez, E.~Peinado, and S.~Ramos-S{\'a}nchez, JHEP \textbf{12}
  (2016), 131, \texttt{arXiv:1607.06812} [hep-ph].

\bibitem{Baur:2019kwi}
A.~Baur, H.~P. Nilles, A.~Trautner, and P.~K.~S. Vaudrevange,
  \texttt{arXiv:1901.03251} [hep-th].

\bibitem{Ibanez:1986tp}
L.~E. Ib{\'a}{\~n}ez, H.~P. Nilles, and F.~Quevedo, Phys. Lett. \textbf{B187}
  (1987), 25.

\bibitem{Kobayashi:2004ya}
T.~Kobayashi, S.~Raby, and R.-J. Zhang, Nucl. Phys. \textbf{B704} (2005), 3,
  \texttt{hep-ph/0409098}.

\bibitem{Babu:2002tx}
K.~Babu, I.~Gogoladze, and K.~Wang, Nucl. Phys. \textbf{B660} (2003), 322,
  \texttt{arXiv:hep-ph/0212245} [hep-ph].

\bibitem{Lee:2010gv}
H.~M. Lee, S.~Raby, M.~Ratz, G.~G. Ross, R.~Schieren, et~al., Phys. Lett.
  \textbf{B694} (2011), 491, \texttt{arXiv:1009.0905} [hep-ph].

\bibitem{Lee:2011dya}
H.~M. Lee, S.~Raby, M.~Ratz, G.~G. Ross, R.~Schieren, et~al., Nucl. Phys.
  \textbf{B850} (2011), 1, \texttt{arXiv:1102.3595} [hep-ph].

\bibitem{Kappl:2010yu}
R.~Kappl, B.~Petersen, S.~Raby, M.~Ratz, R.~Schieren, and P.~K. Vaudrevange,
  Nucl. Phys. \textbf{B847} (2011), 325, \texttt{arXiv:1012.4574} [hep-th].

\end{thebibliography}
\bibliographystyle{NewArXiv}

\end{document}